\documentclass[final,5p,times,twocolumn]{elsarticle}

\usepackage{booktabs}
\usepackage{tabularx}
\usepackage{float}
\usepackage{color,soul}
\usepackage{amssymb}

\usepackage{amsmath,array,graphicx}

\usepackage{pdfcolparallel}

\makeatletter
\def\@author#1{\g@addto@macro\elsauthors{\normalsize%
    \def\baselinestretch{1}%
    \upshape\authorsep#1\unskip\textsuperscript{%
      \ifx\@fnmark\@empty\else\unskip\sep\@fnmark\let\sep=,\fi
      \ifx\@corref\@empty\else\unskip\sep\@corref\let\sep=,\fi
      }%
    \def\authorsep{\unskip,\space}%
    \global\let\@fnmark\@empty
    \global\let\@corref\@empty  
    \global\let\sep\@empty}%
    \@eadauthor={#1}
}
\makeatother

\usepackage[table]{xcolor}
\definecolor{mycolor}{RGB}{255,255,204}

\usepackage{lineno}				

\usepackage{hyperref}
\hypersetup{
    colorlinks=true,
    linkcolor=blue,
    filecolor=magenta,      
    urlcolor=cyan,
}
\modulolinenumbers[5]			

\usepackage{subcaption,graphicx}
\usepackage{makecell}

\journal{Astronomy and Computing}

\biboptions{authoryear}		

\begin{document}

\begin{frontmatter}

\title{Using the Agile Software Development Lifecycle to develop a standalone application for generating Colour Magnitude Diagrams}



\author[add1,add2]{K. Fitzgerald\fnref{fn1}}
\fntext[fn1]{Email: Mr. Karol Fitzgerald - k.fitzgerald12@nuigalway.ie}
	
\author[add1]{L-M. Browne\fnref{fn2}}
\fntext[fn2]{Email: Ms. Lisa-Marie Bowne - l.browne1@nuigalway.ie}

\author[add1]{R.F. Butler\fnref{fn3}\corref{cor1}}
\cortext[cor1]{Corresponding author}
\fntext[fn3]{Email: Dr. Ray Butler - ray.butler@nuigalway.ie}

\address[add1]{Centre for Astronomy, School of Physics, National University of Ireland - Galway, University Road, Galway, H91 CF50, Ireland}
\address[add2]{Permanent address: Athlone Institute of Technology, Dublin Road, Bunnavally, Athlone, Co. Westmeath, N37 HD68, Ireland}
\address[add3]{Declarations of Interest: none for any author}




\begin{abstract}
Virtual observatories allow the means by which an astronomer is able to discover, access, and process data seamlessly, regardless of its physical location.  However, steep learning curves are often required to become proficient in the software employed to access, analyse and visualise this trove of data.  It would be desirable, for both research and educational purposes, to have applications which allow users to visualise data at the click of a button. Therefore, we have developed a standalone application (written in Python) for plotting photometric Colour Magnitude Diagrams (CMDs) - one of the most widely used tools for studying and teaching about astronomical populations.  The  \textit{CMD Plot Tool} application functions \textit{``out of the box"} without the need for the user to install code interpreters, additional libraries and modules, or to modify system paths; and it is available on multiple platforms.  Interacting via a graphical user interface (GUI), users can quickly and easily generate high quality plots, annotated and labelled as desired, from various data sources. This paper describes how \textit{CMD Plot Tool} was developed using Object Orientated Programming and a formal software design lifecycle (SDLC). We highlight the need for the astronomical software development culture to identify appropriate programming paradigms and SDLCs. We outline the functionality and uses of \textit{CMD Plot Tool}, with examples of star cluster photometry.  All results plots were created using \textit{CMD Plot Tool} on data readily available from various online virtual observatories, or acquired from observations and reduced with IRAF/PyRAF.
\end{abstract}


\end{frontmatter}



\section{Introduction} In recent years, online archives and virtual observatories or VOs \citep{2015_virtual_observatory} have increased the availability and quantity of digitised, multi-wavelength astronomical data accessible to research scientists, academics, students and hobbyist astronomers. As the data access problem diminishes, more focus is now required on effectively and efficiently analysing, interpreting and visualising results from the data. To one major institution active in the VO sphere, the European Southern Observatory, the term "Virtual Observatory Tools" only means tools to query and retrieve raw or processed data from its archives\footnote{\url{https://archive.eso.org/cms/virtual-observatory-tools.html}}; not to manipulate or visualise it. Popular examples of existing tools which do address some forms of manipulation and visualisation include SAO ds9 (available as a desktop application), which is restricted to pixel data and has limited options for overplotting of markers and regions; CDS Aladin (available in both full-functionality desktop and "Lite"-functionality HTML5 browser versions), which can perform many operations on VO images, catalogs and overlays - but lacks the ability to plot an "x vs y" graph such as a CMD; and TOPCAT and VOPlot, both of which operate on VO catalog (tabular) data. EURO-VO maintains an online list\footnote{\url{http://www.euro-vo.org/?q=science/software}} of these and other tools for VO data. But what this list makes clear is that no one piece of user-friendly software is readily available to manage, analyse and visualise this diversity of data.  As a result of this, a growing number of researchers, students and hobbyists are developing their own tools for their own specific purposes.

\begin{figure}[!t]
	\centering
	{\includegraphics[width=3.5in]{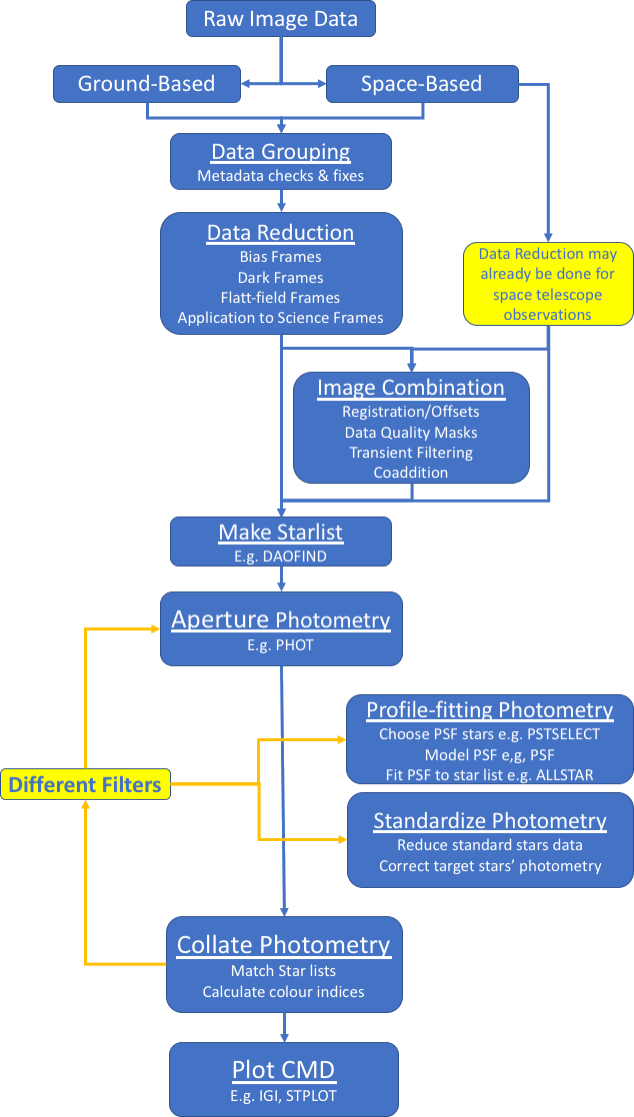}} %
	\caption[CMD Generation Flowchart]{CMD Generation Flowchart. This illustrates the steps and learning curve required to generate a CMD of stellar objects, starting with raw image data. Task names in IRAF/PyRAF/DAOPHOT, which would achieve the photometry steps, are given for illustration.}
	\label{MakePlotDAO}
\end{figure}

Furthermore, new users also experience steep learning curves in installing, understanding and utilising specialist software. One example is where a user must create a colour magnitude diagram (CMD) to visualise and interpret photometric data. A CMD is an observational representation of the Hertzsprung-Russell diagram \citep{1913_Russell}, not in physical units (absolute luminosity and temperature) but in observational units (stellar magnitudes and magnitude differences, in particular filter bandpasses). The flowchart shown in Figure \ref{MakePlotDAO} outlines the steps involved in generating a CMD using IRAF\footnote{IRAF: Image Reduction and Analysis Facility.} (ascl:9911.002) \citep{1993_IRAF} or PyRAF\footnote{PyRAF is a product of the Space Telescope Science Institute, which is operated by AURA for NASA.} (ascl:1207.011) \citep{2000_PyRAF}. The sophistication, flexibility and breadth of IRAF have made it one of the ``industry standard" platforms in observational astronomy research - but is it necessary or reasonable to expect mastery in IRAF (or its ilk) as a pre-requisite for CMD generation? Even tools like VOPlot and TOPCAT, which take photometric catalog data as their starting point, may intimidate some users: VOPlot has a 64-page online user manual; TOPCAT's PDF manual runs to 527 pages! This paper introduces \textit{CMD Plot Tool}: a standalone application designed to read various sources of data and to easily produce consistent and high quality plots of CMDs - allowing novices to bypass the usual learning curve, and experts to get results more readily. The executables and source code are publically available on Sourceforge at \url{https://cmd-plot-tool.sourceforge.io}

The paper outline is as follows: Section 2 discusses the development of specialised software in astronomy and recent research highlighting its importance and current trends.  The software development lifecycle (SDLC) used in developing \textit{CMD Plot Tool}, and the process of software freezing to create an \textit{``out of the box"} application, are outlined.  These techniques have wider utility for other application developers. This section concludes by documenting how this application can be deployed on various platforms without the need for additional development environments, code interpreters, or users having to modify system paths.  Section \ref{CMDfunctionality} describes the functionality of the \textit{CMD Plot Tool}, how the various data sources are integrated, and the operations needed to produce CMDs.  Section \ref{cmdSection} illustrates how CMDs can be used to interpret, visualise and analyse celestial populations, with focus on sample datasets to produce CMDs of globular clusters (GCs).    Section \ref{theConclusion} presents our conclusions.

\section{Software Development in Astronomy}
A  survey carried out by \cite{2015_Astro_Software_Survey} focused on the use of software and the software skills of the participants in the worldwide astronomy community, between December 2014 and February 2015.  Participants consisted of 1142 astronomers: 380 graduate students, 340 postdocs, 385 research scientists and faculty, and the remaining 37 consisted of undergraduate students, observatory scientists, etc.  All survey participants responded ``yes" when asked if they used software as part of their research; 11\% of responders said they used software developed by others; 57\% used software developed by themselves and others; while 33\% said that they used software they developed themselves for specific purposes, as there was no software readily available.  This research also revealed the open source language, Python, to be the programming language of choice.   However, there was no reference in this research to the software development life cycle (SDLC) employed by developers in the astronomy domain.

Prior to our development of \textit{CMD Plot Tool}, we investigated which (if any) SDLCs are utilised in the development of astronomy-specific software. Examples of promising software publications we checked include 
\citet{2017_AstroImageJ}, \citet{2015_Starfish}, \citet{2014_Software_Big_Data}, \citet{2012_AstroML}, \citet{2014_Dogfood_Data}, \citet{2015_Software_Registry}, \citet{2014_Astropy_BuildingBlocks}, \citet{2013_KMOS_Reduction_Software}, \citet{2013_Blender}, \citet{2012_Visual_High_Data}, and \citet{2007_CASA}. We found no information regarding the process (programming paradigms) by which the published astronomy-specific software was written, i.e. the developers do not state whether they employed any particular SDLC or test driven development (TDD: \citet{beck2002driven, TDD_Astels_2003}) approach. In order to preclude our own selection bias, we then checked \textit{every} paper in the most recent year's volume of freely available Astronomical Data Analysis Software and Systems (ADASS) conference e-proceedings \citep{2015_ADASS} - choosing ADASS because of its pre-eminence as a forum for communicating best practice in this field. Not all of these 128 ADASS papers involve writing code - in fact, over half are concerned with other topics (databases, data standards and models, metadata and archive management, VO interfacing, creative use of archival data, collaborative workspaces, cloud computing, source code libraries). However, our trawl revealed that in nearly all cases where a code or pipeline was developed (typically using Python, Java, C or C++; sometimes in conjunction with MPI or Perl), it is described in terms of its algorithms and functionality, layers and internal architecture, external connectivity, inputs and outputs; but not in terms of its development approach. The few exceptions to this rule were \cite{ESO_HLDRL}, \cite{Vallejo_BepiColombo}, \cite{PalomarTF} and \cite{MeerkatTouch}. The first of these papers outlines a process of iterating with users on their requirements, but does not explicitly identify this approach by its name. The other three, in contrast, give comprehensive treatment to their Agile and User-Centred design methodologies. This investigation indicates that only the order of 5$\%$  of astronomical coding papers provide such information, and presumably this reflects the proportion of projects which take a SDLC approach.

The advantages of employing and adhering to a SDLC include: clarity in project objectives, requirements and estimates; more stable systems where missing functionality can be easily identified; developing a valuable relationship between users and developers.  A lack of awareness of these benefits is an unfortunate aspect of the astronomical software development culture, and if unchecked can potentially lead to software/systems being over budget, delivered late, or missing functionality; and in worst cases, to complete project failure.  So in an attempt to steer this culture, we will describe programming paradigms and the SDLC/TDD approach, including the advantages and disadvantages of various SDLCs.

\subsection{Software Development Life Cycle (SDLC) and Programming Paradigms}
A SDLC  defines a structured sequence of stages in software engineering to develop the intended software product.  A TDD approach relies on a shorter development cycle, where requirements become specific test cases that the software must pass.  Another factor to consider is determining which programming paradigm to use. In developing \textit{CMD Plot Tool}, much emphasis was put on determining an appropriate SDLC/TDD approach and programming paradigm, while also setting out a strict set of requirements beforehand.  This section will discuss the options available and justify the choices made. 

\paragraph{Programming Paradigms}   
Different paradigms allow for alternative approaches when developing software applications. However, it is important to note that while programming languages are usually classified by one paradigm, there are some languages - such as Python - that can handle multiple paradigms.  The two most common programming paradigms are the  \textit{procedural paradigm} and the  \textit{object-oriented paradigm}.  Procedural programming relies on the premise that the coder utilises procedures (routines or subroutines) to operate and manipulate data.  This type of programming is sequential in nature, and so not particularly complicated.  Object-orientated programming (OOP) amalgamates procedures and data into \textit{objects}, allowing for more complicated functionality, while minimising the amount of code required. Objects can either be independent or associated with other objects, and they interact by passing information to each other.  In the instance where an object interacts with another object, regardless of their similarity or differences, then the object contains information about itself (\textit{encapsulation}) and the objects it can interact with (\textit{inheritance}).  Inheritance enables new objects to inherit the properties and methods of existing objects. OOP also utilises \textit{classes}, which are user-defined prototypes for an object.  They define a set of attributes that characterise any object of the class.  Classes allow for the generation of multiple objects of the same type that can be used anywhere in the code, allowing for a significant reduction in coding but also more flexibility and functionality when dealing with multiple objects of the same type.  Inheritance may be exploited for further coding efficiencies: rather than creating a new object from scratch, developers can reference a pre-existing object or superclass and create a subclass based on the previous one, allowing them to reuse code and functionality more effectively.  

Using OOP, developers can manage and break software projects down into smaller, more manageable modular problems, one object at a time.  The modularity of objects makes trouble-shooting easier; encapsulation ensures that objects are self-contained and functionality contained within methods/functions specific to that class.  When errors do occur, developers know where to look without having to navigate through large amounts of code.  Another advantage of OOP is in software maintenance; applications may evolve by additional functionality or improvements to the user experience, and upgrades might be required to allow software to work on newer computer systems.  While an OOP-based application requires a great deal of planning pre-release, less work is needed to maintain it over time.  Given these advantages, \textit{CMD Plot Tool} was developed using OOP, written in Python.

\begin{figure}[!h]
	\centering
	{\includegraphics[width=3.5in]{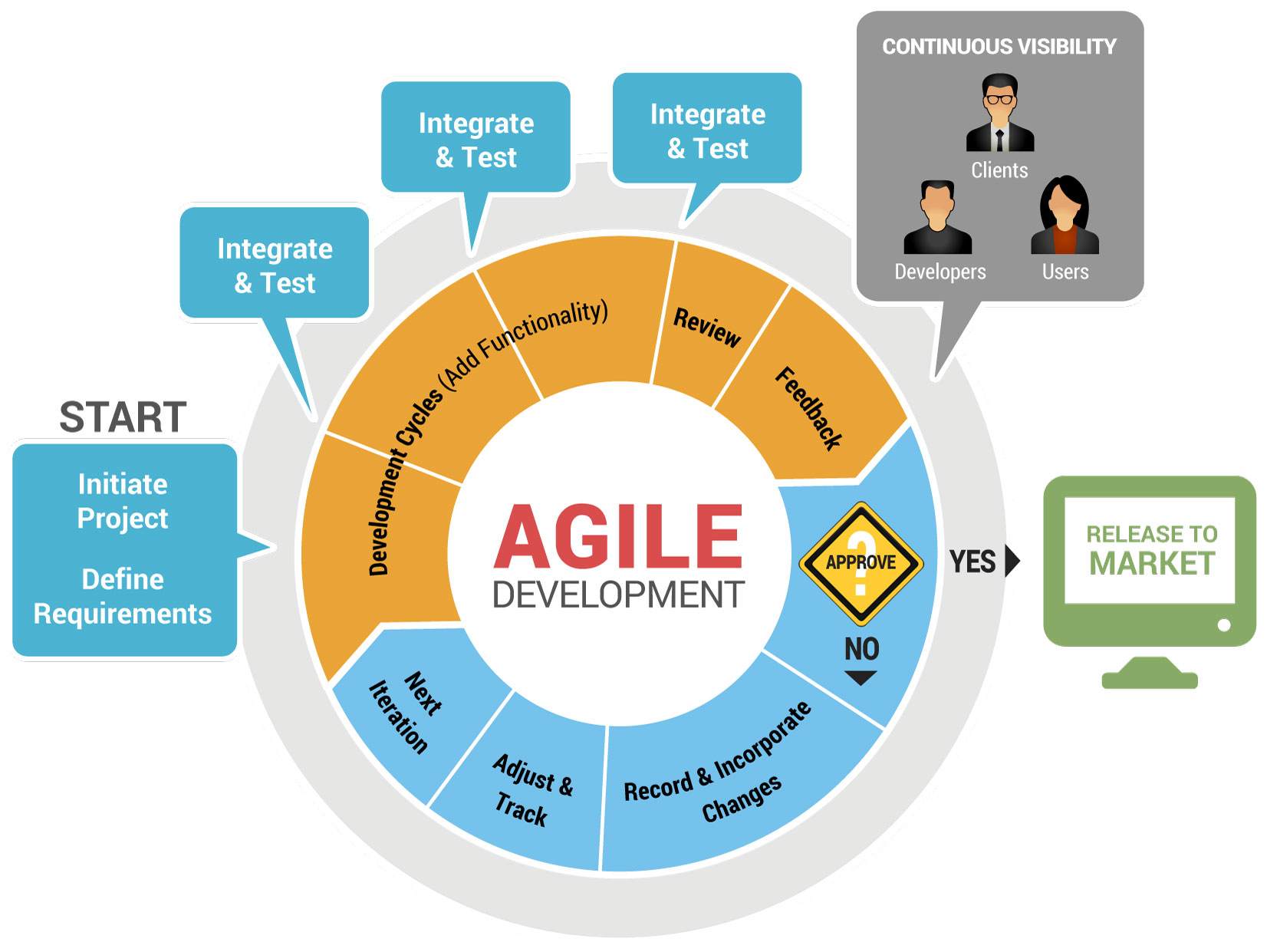}} %
	\caption[Agile Software Lifecycle]{The Agile Software Development Lifecycle. \textit{Image reproduced from \url{http://cssmith.co/wp-content/uploads/2017/10/scrum-diagram-agile-application-and-cloud-computing.jpg}.}}
	\label{agile}
\end{figure}

\paragraph{Agile Software Development}
The Agile Software Development Lifecycle (ASDL) is a set of methodologies (four values and twelve principles) defined by the Agile Manifesto \citep{AgileWEB}.  It values:

\begin{enumerate}
\item{\textbf{Individuals and interactions} over processes and tools}
\item{\textbf{Working software} over comprehensive documentation}
\item{\textbf{Customer collaboration} over contract negotiation}
\item{\textbf{Responding to change} over following a plan}
\end{enumerate}

The twelve principles expand on these values in more detail, giving more emphasis to the items highlighted in bold. The ASDL is illustrated in Figure \ref{agile}. It combines iterative and incremental process models, and focuses on adaptability and satisfaction by rapid delivery of working software products before the product is reviewed.   During each development lifecycle, additional functionality is incorporated and tested.  If the product meets the specification and satisfaction of the project owner, then the product is released to the market or end users; if not, then another iteration of the development phase begins after all incremental changes have been noted.   Each development phase focuses on particular functionality.

\begin{table*}[htbp]
\centering
\begin{tabularx}{\linewidth}{| l| l | X |}
\hline
 & \textbf{Agile} & \textbf{Traditional}  \tabularnewline  \hline
User requirement & Iterative acquisition & Detailed user requirements are well defined before coding \tabularnewline  \hline
Development direction & Readily changeable & Fixed \tabularnewline  \hline
Testing & On every iteration & After coding phase completed \tabularnewline  \hline
Customer involvement & High & Low \tabularnewline  \hline
Extra quality required for developers & Interpersonal skills \& basic business knowledge & Nothing in particular \tabularnewline  \hline
Suitable project scale & Low to medium scaled & Large-scaled \tabularnewline  \hline
\end{tabularx}
\caption{Agile SDLC versus Traditional SDLC models \citep{2012Agile_Trad1}}
\label{AvT}   
\end{table*}

\begin{table*}[ht!]	
	\centering
	\begin{tabularx}{\linewidth}{| l| l | l | l| l | X | l | }
	\hline
\textbf{Factors} & \textbf{Waterfall} & \textbf{V-Shaped} & \textbf{Evolutionary Prototyping} & \textbf{Spiral} & \textbf{Iterative \& Incremental} & \textbf{Agile} \tabularnewline  \hline	
Unclear User Requirement & Poor & Poor & Good & Excellent & Good & Excellent \tabularnewline  \hline	
Unfamiliar Technology & Poor & Poor & Excellent & Excellent & Good & Poor \tabularnewline  \hline	
Complex System & Good & Good & Excellent & Excellent & Good & Poor \tabularnewline  \hline	
Reliable system & Good & Good & Poor & Excellent & Good & Good \tabularnewline  \hline	
Short Time Schedule & Poor & Poor & Good & Poor & Excellent & Excellent \tabularnewline  \hline	
Strong Project Management & Excellent & Excellent & Excellent & Excellent & Excellent & Excellent \tabularnewline  \hline	
Cost limitation & Poor & Poor & Poor & Poor & Excellent & Excellent \tabularnewline  \hline	
Visibility of Stakeholders & Good & Good & Excellent & Excellent & Good & Excellent \tabularnewline  \hline	
Skills limitation & Good & Good & Poor & Poor & Good & Poor \tabularnewline  \hline	
Documentation & Excellent & Excellent & Good & Good & Excellent & Poor \tabularnewline  \hline	
Component reusability & Excellent & Excellent & Poor & Poor & Excellent & Poor \tabularnewline  \hline	
\end{tabularx}	
	\caption{Rubric used in deciding on the SDL for this project \citep{AgileRubric}}
	\label{table:rubric}   
\end{table*}


A comparison of the ASDL and traditional software development life cycles is presented in Table \ref{AvT} \citep{2012Agile_Trad1}. This explains our main reasoning for choosing the ASDL over traditional SDL:  the project (1) was relatively low scale, (2) had high customer involvement, and (3) was characterised by a changeable development direction and evolving user requirements.  Table \ref{table:rubric} \citep{AgileRubric} outlines the rubric used in deciding on the ASDL over the alternatives for this project; ASDL again scores highest or joint-highest in factors such as user requirement and ``visibility of stakeholders" (another name for customer involvement), but low cost and short time schedule were also important in a busy, non-commercial, academic development team.

Our utilisation of the ASDL to develop of \textit{CMD Plot Tool} proved extremely beneficial.  \textit{User stories} - concise, written descriptions of specific functionality that is valued to the user or owner of the software - were generated by two co-authors (R.B., L.-M.B.) prior to the commencement of coding by K.F. A description of a user story follows the following template \citep{UserStories}:

\begin{center}
	\textit{As a \textbf{[user role]}, I want to \textbf{[goal]}, so i can \textbf{[reason]}} \\
\end{center}

For example: \textit{As an astronomer, I want to be able to select two magnitude files output by IRAF/DAOPHOT; calculate the colour index for each star; format the axis, title, plotting colour; and add annotation; so I can generate a CMD plot for the classroom or for publication.}\\

User stories must be detailed enough to start work; however, further details can be established and clarified as the project progresses.  The relatively small scale of this project benefited the authors by allowing them to discuss the overall goal and functionality of the software prior to and during development.  Regular communication and continuous inputs from the customers (in this instance the co-authors) left little room for guesswork regarding functionality. Requirements specified by the customers were developed into 13 user stories (often with some elements in common).

These user stories focused the development cycle for each iteration of \textit{CMD Plot Tool} and saved time: by allowing the developer to concentrate on breaking requirements down into functionality, this breaks a large project into smaller, more manageable tasks (in total 14 iterations were required, with 1-2 weeks allocated for each depending on the complexity of the functionality).  For example, the above user story is accomplished by the following tasks: 

\begin{enumerate}
\item{Read in DAOPHOT-format photometry file for each filter/waveband}
\item{Extract necessary fields from each file.}
\item{Merge fields from each file, removing errors \& duplicates.}
\item{Calculate the colour indices from pair(s) of magnitudes, and calculate their errors.}
\item{Retrieve user specified parameters regarding axis scale, annotation, and colour options from the GUI.}
\item{Generate and save the plot.}
\item{Display the plot.}
\item{Delete temporary files.}
\end{enumerate}

\begin{figure}[!h]
	\centering
	{\includegraphics[width=3.5in]{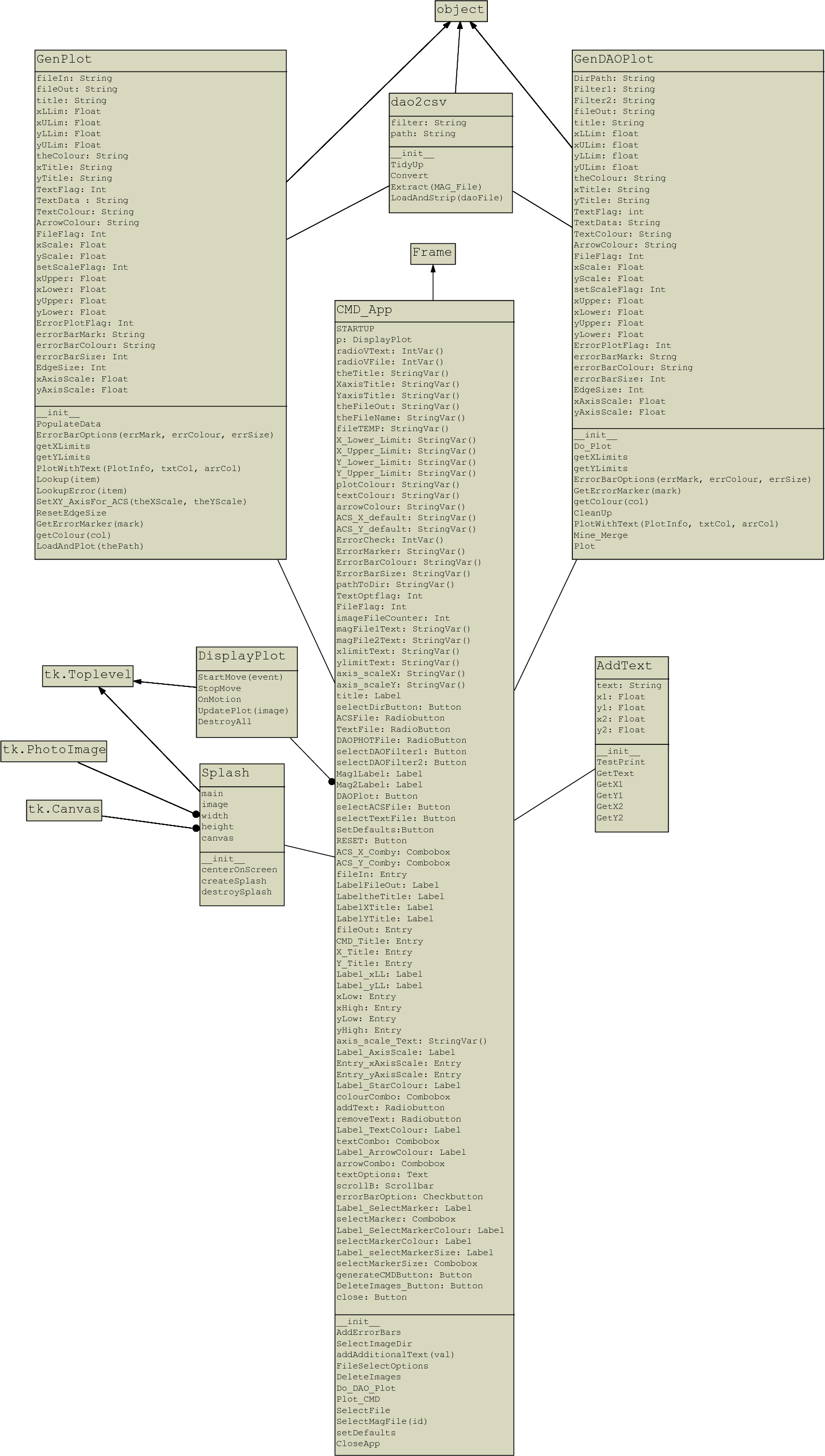}}  
	\caption[UML Diagram]{UML diagram for classes outlined in the Python files in Table \ref{table:code} }
	\label{CMD_UML}
\end{figure}

Another benefit of having well defined user stories  is the ability to anticipate the functionality of each software class and how they interact with each other.  Upon completion, \textit{CMD Plot Tool} consisted of 6 classes, illustrated in the UML diagram of Figure \ref{CMD_UML}, with approximately 1000 lines of code; an analysis of each class is given in Table \ref{table:code}.    

\begin{table*}[htbp]
	\centering
	\begin{tabularx}{\linewidth}{| l | c | X |}
		\hline
		\multicolumn{1}{|l|}{\textit{\textbf{\makecell[l]{Class Name \\ (.py)}}}} & \multicolumn{1}{l|}{\textit{\textbf{\makecell[l]{Lines of\\Code\\(Including\\Comments)}}}} &\multicolumn{1}{l|}{\textit{\textbf{Description}}}  \tabularnewline 
		\hline
SplashScreen  & 41 & Creates and positions a splash screen in the centre of the monitor when the application starts. \tabularnewline  \hline
CMD\_Plot\_Tool & 554 & Constructs and manages operations via the GUI.  Parameters and operational requests based on user interaction are passed to appropriate classes.  \tabularnewline  \hline
Plot\_DAOPHOT   & 189  & Selected DAOPHOT-format photometry (magnitude) files are passed to this class, which constructs and saves a CMD plot based on user-specified parameters. \tabularnewline  \hline
Plotting       & 250  & Selected .zpt and .txt photometry (magnitude) files are passed to this class, which constructs and saves a CMD plot based on the user-specified parameters and requirements.  \tabularnewline  \hline
Plot\_Text      & 21 &  Manages every instance of an annotation on the plot.  A reference to the annotation text, with its corresponding x,y coordinate and its pointer x,y coordinate, allows for multiple instances within a single plot.\tabularnewline \hline
ThePlot        & 41 & Updates and displays the plot on the screen.  Additional functionality also allows for this window to be moved, based on mouse clicks and movement. \tabularnewline  \hline
	\end{tabularx}
	\caption{Characteristics of each class within \textit{CMD Plot Tool}.  Note that the "lines of code" column includes comments, i.e. non-executable code.}
	\label{table:code}   
\end{table*}  

\begin{figure}[!ht]
	\centering
	{\includegraphics[width=2in]{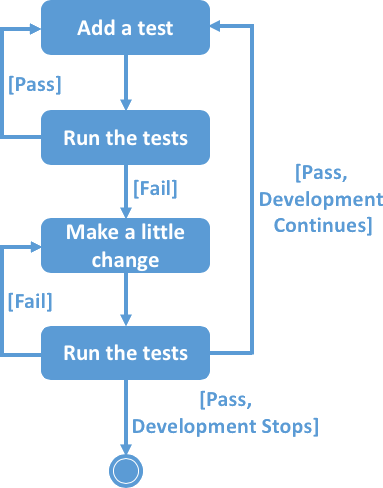}} %
	\caption[TDD Diagram]{Test Driven Development \textit{Image reproduced from \url{http://agiledata.org/essays/tdd.html}.} \textcopyright \, Scott W Ambler}
	\label{TDD_Outline}
\end{figure}

\paragraph{Software Testing} Software testing is essential in evaluating the quality, performance and accuracy of code by detecting differences between actual and expected outputs, for given inputs.  The ASDL (Figure \ref{agile}) recommends that this process be carried out during each development phase, to ensure \textit{verification} and \textit{validation} of the product.  Verification ensures that the product satisfies the conditions imposed at the start of the development phase, to confirm that the application behaves as it should.  Validation ensures the application satisfies the  requirements at the end of the development phase to certify that it is built as per the customer requirements.  There are numerous types of software testing, which can be categorised as being either functional, non-functional or maintenance.  TDD, unit testing and GUI testing were considered as possible testing techniques for testing the \textit{CMD Plot Tool}.

TDD is a technique developed by Kent Beck \citep{beck2002driven}, (Figure \ref{TDD_Outline}) for constructing software. It follows three simple steps repeatedly: 
\begin{itemize}
\setlength\itemsep{-0.5em} 
\item Construct a test for the next functionality to be coded
\item Write functional code until the test is passed
\item Refactor new and old code, making it well structured
\end{itemize}
This particular software development technique relies on the repetition of a very short development cycle where requirements are turned into very specific test cases, and software is then improved to pass only the new tests.  However, implementing a TTD approach in this project proved difficult due to lack of experience in designing test cases prior to coding.  As a consequence of this steep learning curve, it was decided to use unit testing.

The \textit{CMD Plot Tool} was developed and tested using the PyCharm IDE (integrated development environment) 2017 Community Edition, Python 2.7 with later iterations using Python 3.7.  The free community edition of PyCharm offers usage of both testing frameworks\footnote{PyCharm Testing Frameworks: \url{https://www.jetbrains.com/help/pycharm/testing-frameworks.html}} and code analysis tools\footnote{PyCharm Editions Comparison: \url{https://www.jetbrains.com/pycharm/features/editions_comparison_matrix.html}}.  Code inspections detect - and suggest corrections for - compiling errors, code inefficiencies including unreachable code, unused code, non-localised string, unresolved method, memory leaks and even spelling mistakes.  We found the code inspection feature extremely useful in identifying and eliminating potential errors prior to running any unit testing.  In addition to this automated testing, the co-authors behind the user stories used their expertise in astronomical photometry to test the GUI, based on a range of inputs and expected outputs - sampling the 3 types of input file format (see \ref{InputTypes}), and input catalogs of sizes which varied up to approximately $10^{5}$ stars.  Testing covered all of the functionality provided in the GUI (see \ref{CMDfunctionality}), including error bars, plot annotation, axis specification, and side-by-side plotting. Figures \ref{CompareNGC6205}, \ref{NGC3201}, \ref{NGC1851_and_47Tuc}, and \ref{CMD_compare} illustrate some of these features.  Feedback was provided to co-author K.F., who then made the appropriate changes, and the code was tested again.  This process was done at the end of each iteration.

\subsection{Freezing Python Code}
Installation of astronomical software packages can be non-trivial, and a source of frustration.  It is often the case that the end-user installs programming interpreters to execute code, only then to be informed that the software package requires the presence of another to work correctly, and that system paths must be set for referencing additional libraries.  Other issues can arise when there are multiple versions of a particular programming language on a system (e.g. \textit{Python 2} and \textit{Python 3}).  

\textit{Freezing} of code allows for the creation of a single executable file that can be distributed to users.  This application/executable contains all the code and any additional resources required to run the application, and includes the Python interpreter that it was developed on.  The major advantage for distributing applications in this manner is that it will work immediately, without the need for the user to have the required version of Python (or any additional necessary libraries) installed on their system.  A disadvantage of generating a single file is that it will be larger, as all necessary libraries are incorporated.  The increase in file size is acceptable when considering other issues, for example ease of installation, running, and portability to other platforms.  Python freezing tools and platforms supported are listed in Table \ref{table:codeFreezing}.  \textit{CMD Plot Tool} was developed on OSX using Python 3.6, and the following libraries were used and frozen with it: \textit{numpy}, \textit{pandas}, \textit{tkinter}, \textit{PIL}, \textit{csv}, \textit{math}, \textit{matplotlib}, \textit{glob}, \textit{os} and \textit{datetime}.   The size of the frozen output file is 152 MB. 

\begin{table}
	\begin{center}
	\begin{tabular}{ | l | c | c | c |}\hline
  		Tool & Apple OSX & Linux & Microsoft Windows  \\  \hline
		py2app & Yes & No & No\\  \hline
		py2exe & No & No & Yes\\  \hline
		pyInstaller & Yes & Yes & Yes\\ \hline
		cx\_Freeze & Yes & Yes & Yes \\  \hline
	\end{tabular}
	\caption{Python freezing tools and the platforms they support. As this software was developed on OSX, \textit{py2app} was used to successfully freeze the code into a single file.  Further single-file applications/executables were created using \textit{cx\_Freeze} for the platforms listed above. These are available with the source code.}
	\label{table:codeFreezing}   
	\end{center}
\end{table}


\section{Functionality of \textit{CMD Plot Tool}}\label{CMDfunctionality}

\begin{figure*}[th!]
    	\begin{subfigure}[t]{0.5\textwidth}
		\centering
		{\includegraphics[width=3.5in]{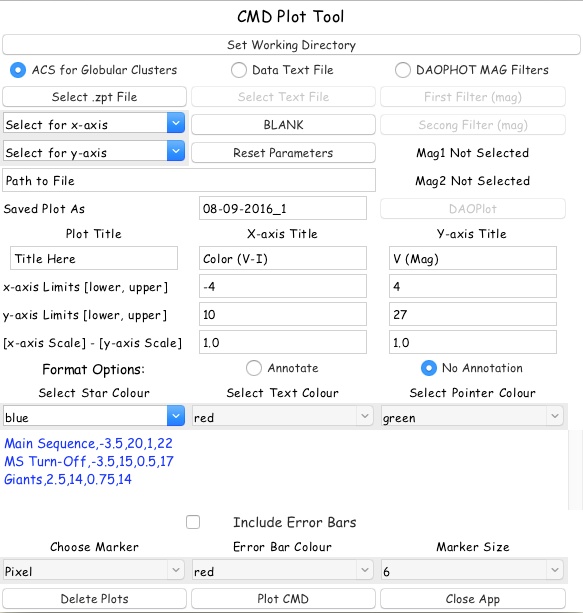}} 
		\caption[CMD App main Window]{}
		\label{CMD_APP}
	\end{subfigure}
    	~ 
	\begin{subfigure}[t]{0.5\textwidth}
        \centering
		{\includegraphics[height=2.25in, width=3.5in]{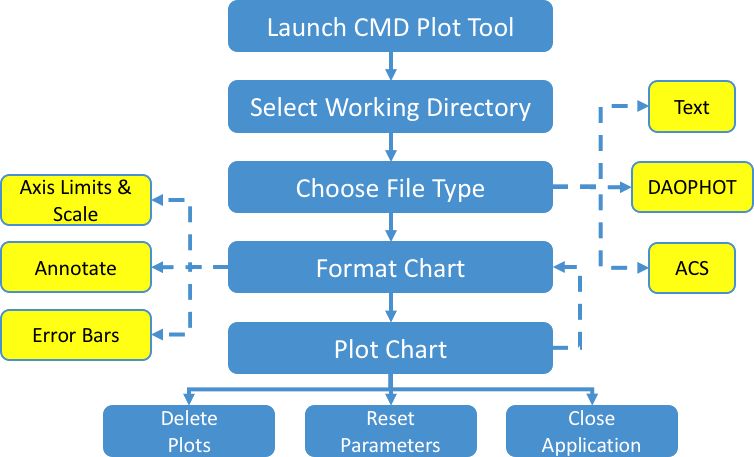}}  
		\caption[Operational Flowchart]{}
		\label{APP_Flow}
	\end{subfigure}
	\caption[\textit{CMD Plot Tool} GUI and Flowchart]{ \textit{CMD Plot Tool}: Panel (a) shows its Graphical User Interface (GUI). Panel (b) outlines its operational flowchart.}
	\label{CMD_APP_and_Flow}
\end{figure*}

\begin{figure*}[!t]
	\centering
	{\includegraphics[width=7in]{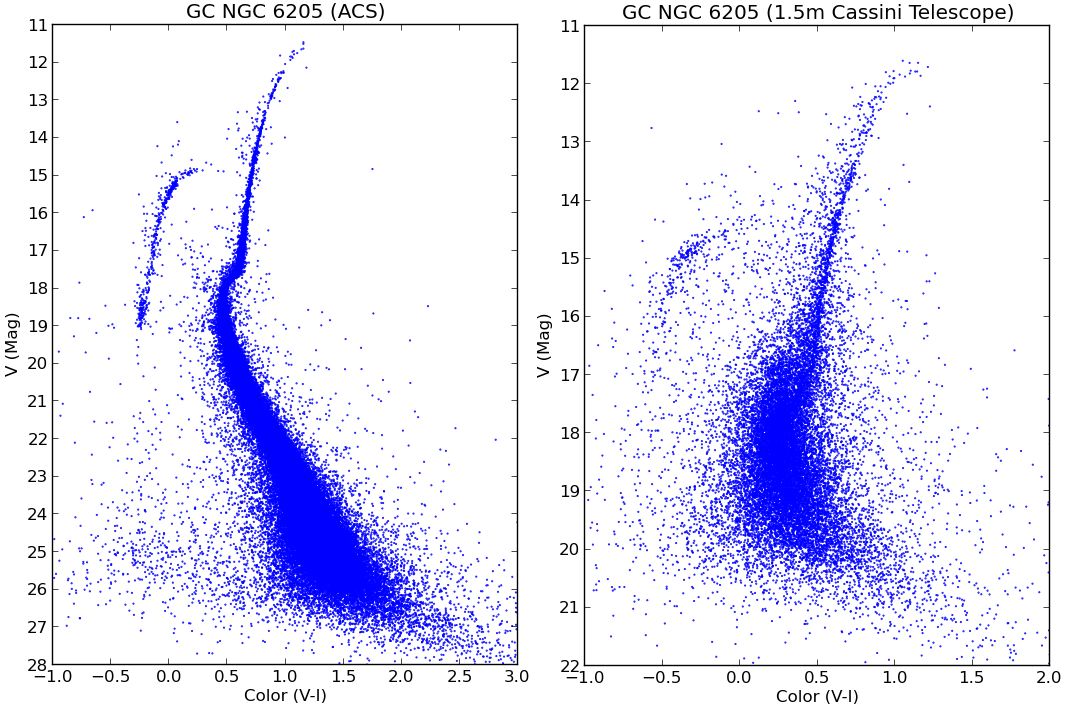}} %
	\caption[Globular Cluster NGC\,6205 Colour Magnitude Diagram]{Comparison of CMDs of globular cluster NGC\,6205 (M13), generated as a single output using the Plot Comparison functionality (see \ref{CMDfunctionality}) of \textit{CMD Plot Tool}.  The plot on the left was generated from reduced Hubble Space Telescope photometry available from the ACS Survey of Galactic Globular Clusters \citep{2007ACS_GC_Survey}.  The plot on the right was generated from data acquired by the authors with the 1.52m Cassini Telescope of the Osservatorio Astronomico di Bologna at Loiano and reduced using IRAF/DAOPHOT.} 
	\label{CompareNGC6205}
\end{figure*}

\begin{figure*}[th!]
    	\begin{subfigure}[t]{0.5\textwidth}
		\centering
		{\includegraphics[width=3.5in]{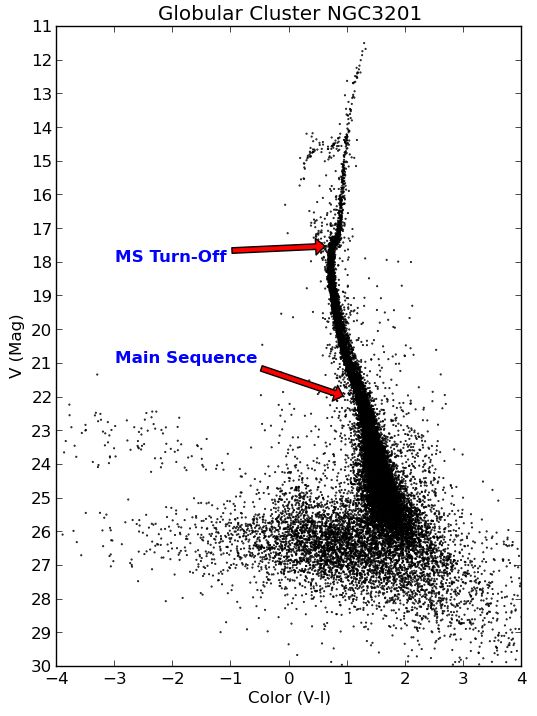}} 
		\caption[Full CMD for GC NGC\,3201]{}
		\label{NGC3201A}
	\end{subfigure}
    	~ 
	\begin{subfigure}[t]{0.5\textwidth}
        \centering
		{\includegraphics[width=3.72in]{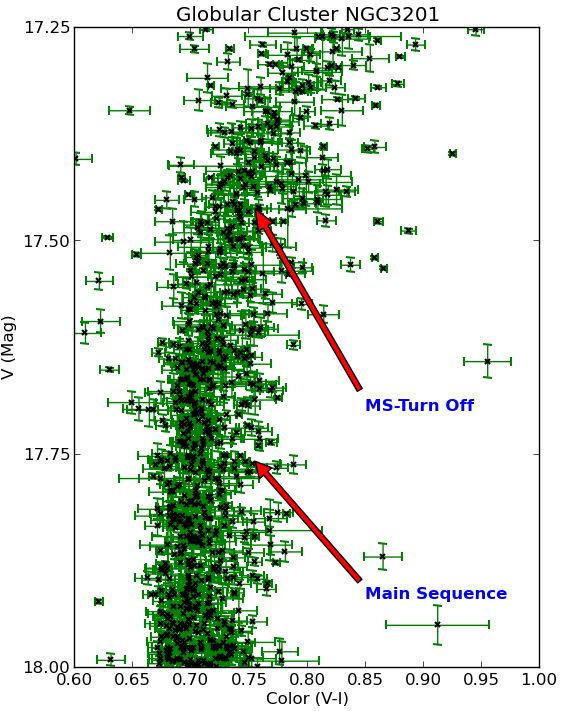}}
	\caption[Zoomed CMD for GC NGC\,3201]{}
	\label{NGC3201B}
	\end{subfigure}
	\caption[CMDs for GC NGC\,3201]{CMDs for GC NGC\,3201, generated from \textit{ACS Survey of Galactic Globular Clusters} data \citep{2007ACS_GC_Survey}. Panel (a) shows the full CMD without error bars, while Panel (b) shows error bars, and is zoomed in and annotated to highlight stars at the top of the main sequence and main sequence turn-off.}
	\label{NGC3201}
\end{figure*}

The Graphical User Interface (GUI) of \textit{CMD Plot Tool}, shown in Figure \ref{CMD_APP}, allows the user to select files and format plots from a single window.  Users generating plots follow the flowchart outlined in Figure \ref{APP_Flow}.  Current functionality includes:
\begin{itemize}
\item{Set working directory: Select the directory where all plots will be saved.}
\item{Save option: Depending on the scenario, users may generate multiple plots.  Each time a plot is generated, it is saved using a timestamp.  This aids efficiency, while also allowing the user to keep a detailed track record of plots generated.  Users can overwrite this option.}
\item{Select input file type: Current functionality allows the user to work on standard text files, \textit{.mag} files generated from IRAF/DAOPHOT, and \textit{.zpt} files from the ACS Survey of Galactic Globular Clusters \citep{2007ACS_GC_Survey}.  Each file type is handled differently due to their unique format. }
\item{Title and Axis options: format plot titles; select which columns in the data are to be plotted on X and Y axes; set axis limit values and units of scale on each axis independently from each other.
 This option enables zooming into particular areas of the CMD plot.  Examples of this are illustrated in Figure \ref{NGC3201B} and Figure \ref{CMD_47Tuc}.}
\item{Annotate options: Allows users to position text and format the colour of text and pointer labels.}
\item{Error bars: Selecting this option displays error bars on each data point, while also allowing the user to format the colour and markers used.  An example of this is illustrated in Figure \ref{NGC3201B}.}
\item{Star count: The user will be informed as to the number of stars plotted within a region.}
\item{Plot comparison: as an educational tool, it is extremely useful to illustrate to students and non-professionals the differences between data acquired from various sources. Figure \ref{CompareNGC6205} is an example of this.}
\item{Delete plots: During the course of plotting CMDs, users may wish to test variations of different parameters.  As a result, and over time, the number of plots generated will increase.  If selected, all plots within the working directory are deleted.}
\item{Reset parameters: Allows the user to reset all parameters within the GUI to their defaults.}
\end{itemize}

\subsection{Supported Input File Types} \label{InputTypes}

\paragraph{Text Files}
The \textit{CMD Plot Tool} offers users the option to plot data contained in standard text files.  The format of the input data depends on the type of plot, i.e. with or without error bars.  Plotting without error bars requires data to be presented in two column format, with each line containing \textit{X Y} paired data, and the first line containing the heading name for each column.  Plotting data with error bars follows the same format, with  additional fields: \textit{X Y X\_error Y\_error}.

\begin{table}[htbp]		
\centering
\begin{tabularx}{\linewidth}{|l|X|}
\hline
\multicolumn{1}{|l|}{\textit{\textbf{Column Name}}} & \multicolumn{1}{l|}{\textit{\textbf{Description}}} \tabularnewline 
\hline
id & Identification of the star  \tabularnewline \hline
x,y & Position of the star on the master image   \tabularnewline \hline
\makecell[l]{Vvega, err,\\VIvega, err, \\Ivega, err} & Magnitudes/colour and errors on the VEGAmag photometric system - stars with one measurement have errors based on root(n) noise. \tabularnewline \hline
Vground, Iground & Magnitudes on the ground VI system.    \tabularnewline \hline
Nv, Ni & Number of times the star was measured in each filter. \tabularnewline \hline
wV, wI & Flags for the nature of image in V and I - short or deep. \tabularnewline \hline
xsig, ysig &  RMS of x and y positions (large scatter indicates poor photometry) \tabularnewline \hline
othv, othi &  Fraction of light from other stars (measure of crowding). \tabularnewline \hline
qfitV, qfitI & Quality of the fits - smaller is better \tabularnewline \hline
RA, Dec & The right ascension and declination (celestial coordinates) of the star at the J2000 equinox, from the HST image headers.\tabularnewline \hline
\end{tabularx}
\caption{ACS \textit{.zpt} file format. Adapted from \href{https://archive.stsci.edu/pub/hlsp/acsggct/hlsp_acsggct_photometry_readme.txt}{\url{https://archive.stsci.edu/pub/hlsp/acsggct/hlsp_acsggct_photometry_readme.txt}} }
\label{ACS_Data}   
\end{table}

\paragraph{ACS \textit{.zpt} Files}	
The  ACS Survey of Galactic Globular Clusters \citep{2007ACS_GC_Survey} is an imaging survey of Galactic GCs using the Advanced Camera for Surveys instrument on board the Hubble Space Telescope (HST).  This HST Treasury project \footnote{Data available online: \url{https://archive.stsci.edu/prepds/acsggct/}} was designed to obtain photometry with S/N $\gtrsim$\,10 for main sequence stars with masses $\gtrsim$\,0.2M$_\odot$ in the central 2 arcminutes of a sample of 66 globulars.  Filters used included F606W ("Wide V" band; yellow light) and F814W ($\sim$I band; near-infrared light). 

The column format of the \textit{.zpt} file is described in Table \ref{ACS_Data}.  Users of \textit{CMD Plot Tool} can select which magnitude/colour columns they wish to plot via a drop down menu that is made available when this file type is selected; normally the VEGAmag columns are used.  Corresponding error bar data is included if the user opts to display error bars on the plot.


\paragraph{DAOPHOT \textit{.mag} files}
DAOPHOT (ascl:1104.011) \citep{1987DAOPHOT} was developed for performing point-source photometry, particularly in crowded stellar fields; initially as a standalone package, and subsequently ported into IRAF.  The typical DAOPHOT \textit{.mag} file (star magnitude file) contains 33 columns of data relating to detected stars, in addition to metadata about the observation and relevant DAOPHOT parameter settings.  Not all of this is necessary for visualising a CMD plot.  \textit{CMD Plot Tool} selects a subset of fields within each \textit{.mag} file (two \textit{.mag} files must be selected for plotting).  Corresponding fields are then merged and sorted based on each star's unique ID within the file.  Calculations are performed based on their unique IDs, magnitudes and magnitude errors.  The results are then saved in temporary storage and plotted.  Once the application is terminated, all temporary files are removed.  

It is important to note that users are not required to have IRAF\,/\,PyRAF installed on their system in order to plot CMDs based on DAOPHOT \textit{.mag} mag files.  The algorithm written to accomplish this utilises the Python Data Analysis Library\footnote{\href{http://pandas.pydata.org}\url{http://pandas.pydata.org}} (\textit{pandas}) and the Python CSV module.  The \textit{pandas} library is an open source library which provides high performance, easy to use data structures (data frames) and other data analytics tools for Python.  Incorporating this with the CSV module allows for fast, reliable and efficient merging and calculating of large scale data.  Performance is excellent even when the dataset being worked on contains tens of thousands of stars.

\begin{table}
\begin{center}
\begin{tabular}{ |l | l | }
\hline
  Colour Options & Marker Options \\ 
 \hline
\cellcolor{blue}\color{white}Blue & Circle ($\circ$) \\  \hline
\cellcolor{red}Red & Triangle ($\bigtriangleup$)\\  \hline
\cellcolor{green}Green & Diamond ($\Diamond$)\\  \hline
\cellcolor{black}\color{white}Black & Point ($\bullet$)\\  \hline
\cellcolor{cyan}Cyan & Pixel ($\cdot$)\\  \hline
\cellcolor{yellow}Yellow & Star ($\star$)\\  \hline
\cellcolor{magenta}\color{white}Magenta & X ($\times$) \\ \hline
\end{tabular}
\caption{Colour options available for text, markers, pointers and error bars, and shape options available for markers.}
\label{options}
\end{center}
\end{table}

\subsection{Annotation}		
The colour options available for text, markers, pointers and error bars, and shape options available for markers, are outlined in Table \ref{options}.  Adding annotation to the plot is done via an algorithm that parses the text input by the user via the text area available on the GUI.  Annotation is entered in the following format: \textit{text, x1, y1, x2, y2}.  \textbf{\textit{Text}} represents the string that the user wishes to display, \textbf{\textit{x1, y1}} represents the start position of the text, and \textbf{\textit{x2, y2}} represents where the labelling arrow will point to.  Data entered is parsed using a comma as a separator, removing unnecessary spaces.  For example, the annotation in Figure \ref{NGC3201B} was entered as follows:

\begin{itemize}
\item[]\textit{Main Sequence, 0.85, 17.92, 0.75, 17.75}\\
\textit{MS-Turn Off, 0.85, 17.70, 0.75, 17.45}
\end{itemize}

\begin{figure*}[th!]
	\centering
    	\begin{subfigure}[t]{0.5\textwidth}
		\centering
	{\includegraphics[width=3.5in]{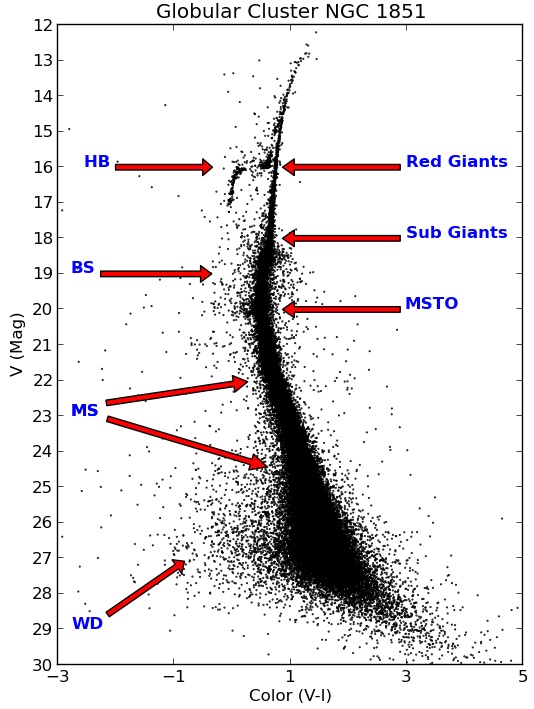}} %
	\caption[CMD of Globular Cluster NGC\,1851]{}
	\label{NGC_1851_B}
	\end{subfigure}%
    	~ 
	\begin{subfigure}[t]{0.5\textwidth}
        \centering
	{\includegraphics[width=3.5in]{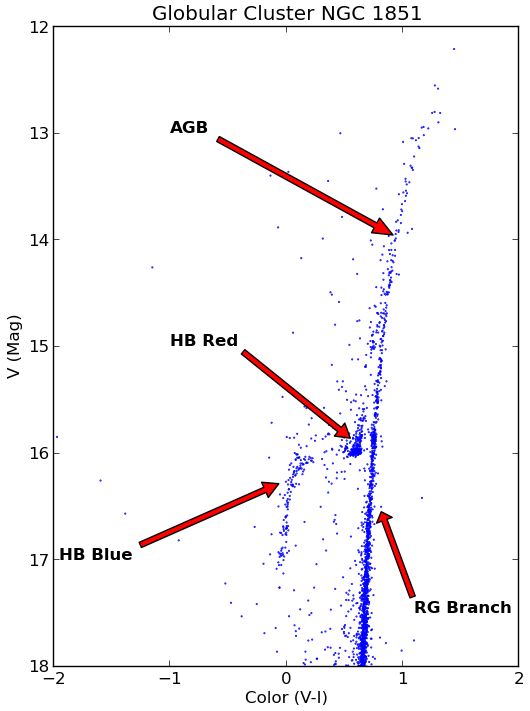}} %
	\caption[Zoomed CMD of Globular Cluster NGC\,1851]{}
	\label{CMD_47Tuc}   
	\end{subfigure}
	\caption[CMDs of Globular Cluster NGC\,1851]{CMDs of Globular Cluster NGC\,1851. Data from the ACS Survey of Galactic Globular Clusters \citep{2007ACS_GC_Survey}. Panel (a) shows the full CMD, while Panel (b) zooms in on the brighter sequences. Explanation of labels:  MS: Main Sequence, MSTO: Main Sequence Turn Off, HB: Horizontal Branch, BS: Blue Stragglers, WD: White Dwarfs, RGB: Red Giant Branch, HB-Blue: blue population of the Horizontal Branch, HB-Red: red population of the Horizontal Branch, AGB: asymptotic giant branch.}
	\label{NGC1851_and_47Tuc}
\end{figure*}

There is no restriction to the amount of annotation on each plot. For example, in Figure \ref{NGC_1851_B} where multiple arrows are projecting from text, the following annotation was applied:
\begin{itemize}
\item[]\textit{MS, -2.75, 23, 0.4, 22}\\
\textit{MS, -2.75, 23, 0.75, 24.5}
%
%
\end{itemize}
If the user wishes to just add text with no pointer, then the \textit{x1, y1} and \textit{x2, y2} coordinates must be the same.

%
%


\section{Usage Case Study for \textit{CMD Plot Tool}: Interpreting the CMDs of Globular Clusters}\label{cmdSection}

\begin{figure*}[!t]
	\centering
	{\includegraphics[width=7in]{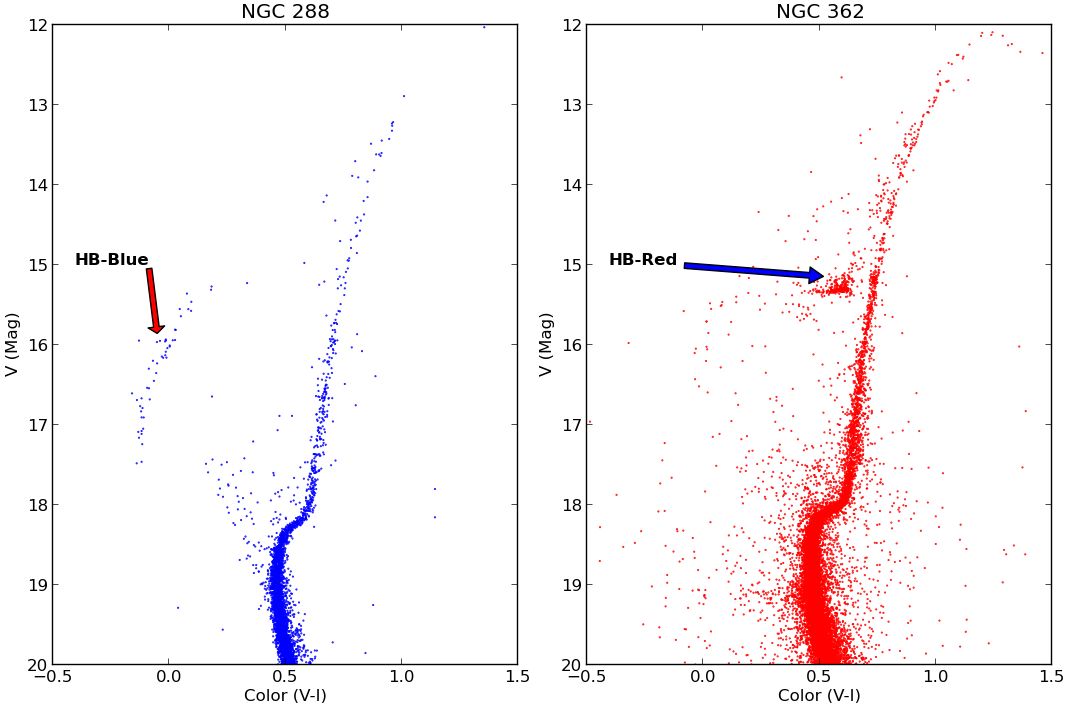}} %
	\caption[Globular Clusters NGC\,288 and NGC\,362 ]{A single output file in Plot Comparison mode. Left side: CMD for GC NGC\,288, where HB-Blue is the blue population of the horizontal branch.  Right side: CMD for GC NGC\,362, where HB-Red is the red population of the horizontal branch. Data from the ACS Survey of Galactic Globular Clusters \citep{2007ACS_GC_Survey}. }
	\label{CMD_compare}   
\end{figure*}

\paragraph{Star Clusters} Stars form in clustered environments which are encapsulated in a giant molecular cloud \citep{2015YoungMassiveCluster}.  They are subjected to the gravitational forces of their stellar brethren, consume fuel to sustain themselves, and transform as they age.  The energy created and exchanged between stars causes gas to heat to a point where it is blown away; during this time some stars may escape a newly forming cluster to become ``runaways".  Those that remain become gravitationally bound.  As a star approaches its end, it will swell to become a red giant; and depending on its mass, it will either eject its bloated atmosphere forming a planetary nebula and exposing its core as white dwarf, or go supernova with a shockwave powerful enough to instigate star formation in other interstellar clouds.  The remains of the core of the exploded star will form a neutron star or black hole; which is created will depend on the mass of the original star.  This lifecycle can take anywhere from a few million to billions of years, again depending on the star.  Star clusters are divided into two principal categories: open (Galactic) clusters and globular clusters.

\paragraph{Globular Clusters}  Globular Clusters (GCs) are among the oldest radiant objects in the universe, with estimated ages between 10\,Gyr $-$ 16\,Gyr \citep{Krauss2003}.  There are approximately 150 known GCs in the Milky Way galaxy, most residing in its spherical halo. Their stellar populations are extremely dense reaching up to ${10^6}$ stars $p{c^{ - 3}}$.  Due to their density, it is highly probable that collisions occur between stars within the cluster, such collisions would lead to a variety of exotic stellar objects such as pulsars, Blue Straggler stars, close interacting binaries, and X-ray sources \citep{M15_FastCam}. GCs are also readily observable in external galaxies due to their compact sizes\footnote{half-light radii of a few parsecs} and masses ranging from ${10^4}\,{M_\odot} - {10^6}\,{M_ \odot}$ \citep{2006_brodie_annual}, although it is harder to resolve extragalactic GCs into individual stars.  

GCs are of interest because of their age, assumed relatively homogeneous populations, relative isolation in their parent galaxies, and abundance of unusual objects, such as blue stragglers, X-ray binaries, radio pulsars, and intermediate-mass black holes \citep{2010YoungMassiveCluster}.  \cite{Krauss2003} describe three independent ways to reliably infer the ages of the oldest star's in our galaxy: (i) Radioactive Dating, (ii) White Dwarf Cooling and (iii) Main Sequence Turnoff Time Scale: a robust prediction of the theoretical models for determining absolute GC age is the time it takes a star to exhaust the supply of hydrogen in its core and leaves the main sequence.  However, uncertainties in determining distances to GCs are a source of doubt when estimating ages.

\subsection{CMDs for Globular Clusters}
The CMD is an extremely useful tool in aiding the understanding of the stellar populations within a cluster, and their evolution. Since a CMD plots apparent magnitudes, stars of the same inherent luminosity, but different distances from the observer, are displaced vertically on a CMD. But since the depth of a star cluster is much less than its distance, all its stars can be taken to be at the same distance.  The colours of the stars, calculated as the ratio of their fluxes (difference in logarithmic magnitudes) in two filter bandpasses, reflect their surface temperature.  Age and chemical composition differences between stars of a given mass are (mainly) evidenced in temperature (colour) differences, displacing stars horizontally on the CMD. But since the stars in a cluster essentially formed from the same gas nebula at the same time, the assumption of equal age and chemical composition also generally holds.  A typical cluster CMD therefore shows rather tightly defined sequences of stars, with the mass of a star determining where it lands in these sequences. As indicated in  Figure \ref{NGC_1851_B} and Figure \ref{CMD_47Tuc}, a typical CMD for a GC will include a main sequence (MS), main sequence turnoff point (MSTO), subgiant branch (SGB), red giant branch (RGB),  horizontal branch (HB), and asymptotic giant branch (AGB). Blue stragglers (BS) may also be present between the MSTO and HB, and if the photometry reaches deep enough, white dwarfs (WDs) may be detected as well. 

These CMD sequences and branches trace the evolutionary processes in the cluster's stars. The relative numbers of stars in each branch also quantify the relative duration of each evolutionary phase - few stars present in a branch means that they quickly move on to the next phase. 

\paragraph{The main sequence (MS)} Most of a star's life is spent on the main sequence (MS): a phase when hydrogen in the star's core is fused into helium, with temperatures reaching $\sim$\,15 million Kelvin. Energy is transported to the star's surface through radiation and convection. There is difference between the CMD of a globular cluster and that of solar neighbourhood: in the latter, main sequence stars do not reach the horizontal branch. This occurs as such young stars are uncommon in GCs. The age of the cluster is estimated by the main sequence turnoff (MSTO).

\paragraph{The subgiant branch (SGB)} This is a narrow section of the CMD, which joins the main sequence at the turnoff point. A star moves onto the turnoff point when the hydrogen in its core has been depleted and an inert helium core created. Hydrogen will continue burning (fusing) in a shell as the core contracts. There is a small increase in the luminosity of the star as the cooler envelope beyond the shell expands. This is known as the subgiant phase. The position of the SGB moves considerably with age, so the narrowness of this feature on a GC CMD limits the stellar formation period to no more than $\sim$2\% of the cluster's age (Stetson, 1993). The metallicity (relative abundance of elements heavier than helium) of the cluster governs the position of the MS and SGB on the CMD.  Metal-poor clusters tend to have a much flatter SGB than that of metal-rich clusters.

\paragraph{The red giant branch (RGB)} As the envelope of the star expands, the effective temperature decreases and it reaches an adiabatic temperature gradient. This is due to convection within the star's interior and the efficiency of energy transportation to the surface. At this point the star will rise quickly up the RGB. When the star reaches the top of the RGB, its core temperature rises and helium fusion begins, producing carbon and oxygen. This is known as the helium flash.

\paragraph{The horizontal branch (HB)} After the helium flash, energy is being released from helium burning in the star's core plus hydrogen burning continues in the thin shell surrounding the core; as a result the star shrinks, becomes hotter and bluer, and jumps to the horizontal branch. CMDs reveal how the metallicity of the cluster influences the composition of the HB. Metal poor clusters have bluer stars than metal rich clusters. RR Lyrae pulsating variable stars can be found on the HB in the instability strip, which is located between red and blue HB stars. The blue portion of the branch consists of helium burning similar to the hydrogen burning of the MS, at shorter time scales. As the star moves to the red portion of the branch, the helium in its core is depleted and has been completely converted into carbon and oxygen. This evolution occurs rapidly and forms a carbon-oxygen inert core. The populations of blue and red stars on the HB can be seen in Figure \ref{CMD_47Tuc}. Although there is a correlation between HB colour and metallicity, there can also be a spread of HB star colours at any given metallicity. For example NGC\,288 and NGC\,362 have identical metallicities; yet NGC\,288 has a very blue HB and NGC\,362 had a very red HB, as can be seen in Figure \ref{CMD_compare}.  

\paragraph{The asymptotic giant branch (AGB)}  The AGB derives its name from the manner in which the evolutionary track approaches the line of the RGB from the left (see Figure \ref{CMD_47Tuc}).  The inert carbon?oxygen core shrinks and heats up, with both helium and hydrogen burning shells around it. The star then moves up the AGB.  Here the star has a higher temperature than when it was on the RGB.  Stars on this branch will again cool and grow quickly. The core of such a star can be unstable, and its outer envelope can be ejected slowly while the core contracts. This leads to the creation of planetary nebulae with white dwarfs at their centres. 

\paragraph{White dwarfs (WD)} Because of their faintness, the most elusive sequence in most CMDs is the white dwarf cooling sequence. The older WDs have had more time to cool and fade, and the rate of cooling can be modelled, so the faintest detectable WD can be dated relative to the brightest WD. 

\paragraph{CMDs and GC age} The oldest GCs have been of particular interest to researchers as they tried to determine their age.  In doing so, they provided an excellent constraint on determining the age of the universe and subsequently on cosmological models used to describe it \citep{2002_GC_Universe_Age}.  Isochrones are synthetic CMD sequences computed for a grid of different ages and metallicities. Determining the best isochrone fit for stars in the region of the turnoff (MSTO) best estimates the age of the GC.  Analysing the variances of these isochrones between the TO and the start of the subgiant branch is shown to be nearly independent of age and chemical abundances.  More recently, \cite{2013_GC_Age_55} determined the ages of 55 GCs using CMDs and an improved $\Delta V_{TO}^{HB}$ method.  In that study, isochrones were fitted to just the TO portion of the observed CMD where morphology is predicted to nearly independent of age and metallicity; furthermore inferred ages were based on the location of the beginning of the subgiant branch. This shows how several features on a CMD can be used for chronology.


\section{Conclusions}\label{theConclusion}
We have presented a new application, \textit{CMD Plot Tool}, for calculating and plotting CMDs.  It can handle multiple file formats to generate professional and customised plots, without the usual steep learning curve.  Development using Python, OOP and a formal software development lifecycle model (Agile) allowed for the creation of an application that can be deployed on multiple systems. It works \textit{``out of the box''} and does not require any installation of development environments, additional libraries or resetting of system paths.  The tool is available as a single application/executable file, with the source code, on the public open-source repository, Sourceforge.  Sample data is also bundled, to demonstrate its complete functionality to users.  Other functionality within this application is the ability to convert DAOPHOT magnitude files to CSV format. We have illustrated one usage of the application: exploring the CMDs of globular star clusters.

\section*{Funding}		
This work was supported by Athlone Institute of Technology (K.F.) and National University of Ireland - Galway (L.-M.B.)

\section*{References}

\bibliography{mybibfile}

\end{document}